\documentstyle[11pt]{article}

\begin{document}

\author{S. L. Lyakhovich, A. A. Sharapov and K. M. Shekhter}
\title{D=6 massive spinning particle }
\date{{\it Department of Physics, Tomsk State University, Tomsk 634050, Russia\\
\vspace{0.5cm}
The talk presented at the Second Sakharov Conference on Physics,\\
Lebedev Physical Institute, Moscow, May 20-24, 1996}}
\maketitle

\begin{abstract}
The massive spinning particle in six-dimensional Minkowski space is
described as a mechanical system with the configuration space ${\ R}%
^{5,1}\times {\ CP}^3$. The action functional of the model is unambigiously
determined by the requirement of identical (off-shell) conservation for the
phase-space counterparts of three Casimir operators of Poincar\'e
group. The model is shown to be exactly solvable. Canonical quantization of
the model leads to the equations on wave functions which prove to be
equivalent to the relativistic wave equations for the irreducible $6d$
fields.
\end{abstract}

\section{Introduction}

The classical description of the relativistic spinning particles is one of
traditional branches of theoretical physics having a long story \cite
{c1,c2,c3,c4}. By now the several approaches to this problem have been
developed. Most of them are based on the enlargement of the Minkowski space
by extra variables, anticommuting \cite{c5,c6,c7,c8,c9} or commuting \cite
{c10,c11,c12,c13,c16}, responsible for the spin evolution.

In the recent paper \cite{c13}, the new model was proposed for a massive
particle of arbitrary spin in $d=4$ Minkowski space to be a mechanical
system with the configuration space $R^{3,1}\times S^2$, where two sphere $%
S^2$ corresponds to the spinning degrees of freedom. It was shown that
principles underlying the model have simple physical and geometrical origin.
Quantization of the model leads to the unitary massive representations of
the Poincar\'e group. The model allows the direct extension to the case of
higher superspin superparticle \cite{c14} and the generalization to the
anti-de Sitter space \cite{c15}.

Despite the apparent simplicity of model's construction its higher
dimensional generalization is not so evident, and the most crucial point is
the choice of configuration space for spin. In this talk we describe the
massive spinning particle in six-dimensional Minkowski space $R^{5,1}$, that
may be considered as a first step towards the uniform model construction for
all higher dimensions. It should be also noted that this generalization may
have a certain interest in its own rights since six is the one in every four
remarkable dimensions: 3, 4, 6 and 10 where the classical theory of
Green-Schwarz superstring can be formulated.

Let us now sketch the broad outlines of the construction. First of all, for
any even dimension $d$, the model's configuration space is chosen to be the
direct product of Minkowski space $R^{d-1,1}$ and some $m$-dimensional
compact manifold $K^m$ being a homogeneous transformation space for the
Lorentz group $SO(d-1,1)$. Then the manifold $M^{d+m}=R^{d-1,1}\times K^m$
proves to be the homogeneous transformation space for Poincar\'e group. The
action of the Poincar\'e group on $M^{d+m}$ is unambiguously lifted up to
the action on the cotangent bundle $T^{*}(M^{d+m})$ being the extended phase
space of the model. It is well-known that the massive unitary irreducible
representations of the Poincar\'e group are uniquely characterized by the
eigenvalues of $d/2$ Casimir operators%
$$
C_1={\bf P}^2\ ,\ C_{i+1}={\bf W}^{A_1...A_{2i-1}}{\bf W}_{A_1...A_{2i-1}}\
,\quad i=1,...,\frac{d-2}2\ ,
$$
where{\bf \ }${\bf W}_{A_1....A_{2i-1}}=\epsilon _{A_1...A_d}{\bf J}%
^{A_{2i}A_{2i+1}}...{\bf J}^{A_{d-2}A_{d-1}}{\bf P}^{A_d}$ and ${\bf J}_{AB},%
{\bf P}_C$ are the Poincar\'e generators. This leads us to require the
identical (off-shell) conservation for the quantum numbers associated with
the phase space counterparts of Casimir operators. In other words $d/2$
first-class constraints should appear in the theory.

Finally, the dimensionality $m$ of the manifold $K^m$ is specified from the
condition that the reduced (physical) phase space of the model should be a
homogeneous symplectic manifold of Poincar\'e group (in fact it should
coincide with coadjoint orbit of maximal dimension $d^2/2$). The simple
calculation leads to $m=d(d-2)/4$. In the case of four-dimensional Minkowski
space this yields $m=2$ and two-sphere $S^2$ turns out to be the unique
candidate for the internal space of the spinning degrees of freedom. In the
case considered in this paper $d=6$, and hence $m=6$. As will be shown below
the suggestive choice for $K^6$ is the complex projective space $CP^3$.

The models can be covariantly quantized \`a la Dirac by imposing the
first-class constraints on the physical states being the smooth complex
functions on the homogeneous space $M^{d(d+2)/4}=R^{d-1,1}\times
K^{d(d-2)/4} $%
$$
(\widehat{C}_i-\delta _i)\Psi =0\quad ,\qquad i=1,...,\frac d2\ ,
$$
where the parameters $\delta _i$ are the quantum numbers characterizing the
massive unitary representation of the Poincar\'e group. Thus the
quantization of the spinning particle theories reduces to the standard
mathematical problem of harmonic analysis on homogeneous spaces. It should
be remarked that manifold $M^{d(d+2)/4}$ may be thought of as the ${\it %
minimal}$ (in sense of its dimensionality) one admitting a non-trivial
dynamics of arbitrary spin, and hence it is natural to expect that the
corresponding Hilbert space of physical states will carry the ${\it %
irreducible}$ representation of the Poincar\'e group.

Our spinor notations and conventions mainly coincide with those adopted in
ref. \cite{c17} except for inverse signature of the Lorentz metric.

\section{Classical theory}

We start to construct the model with describing of the covariant
parametrization for the spinning sector of the configuration space chosen as
$CP^3$. It is useful to realize $CP^3$ as complex projective space
parametrized by the four-component left-handed Weyl spinor $\lambda _{a,}$ $%
a=1,...,4$ subject to the equivalence relation $\lambda _a\sim \alpha
\lambda _a,\ \alpha \in C\backslash \{0\}$. Since the Lorentz
transformations of the spinors obviously commute with the projective ones
generated by the vector fields
\begin{equation}
\label{a1}d=\lambda _a\partial ^a\quad ,\qquad \overline{d}=\overline{%
\lambda }_a\overline{\partial }^a
\end{equation}
the action of the Lorentz group can thereby be transferred from $C^4$ to $%
CP^3$. (In rel. (\ref{a1}) $\overline{\lambda }_a=B_a\!^{\stackrel{.}{a}}%
\overline{\lambda }_{\stackrel{.}{a}}$ and $B$ is the Lorentz invariant
matrix converting the dotted indices into the undotted ones \cite{c17}.) The
action of the Poincar\'e group on $M^{12}=R^{5,1}\times CP^3$ is generated
by the vector fields
\begin{equation}
\label{a2}P_A=\partial _A\quad ,\qquad J_{AB}=x_A\partial _B-x_B\partial
_A+\left( \left( \sigma _{AB}\right) _a\!^b\lambda _b\partial ^a+c.c.\right)
\ ,
\end{equation}
$x^A$ being the Cartesian coordinates on $R^{5,1}$.

To construct the Lagrangian, we consider all the possible Poincar\'e
invariants of the particle's world-line on $M^{12}$. There are only three
first-order invariants\footnote{%
Here we left out the invariants transforming by a total derivative under the
projective transformations which nevertheless may play the crucial role in
attempting to construct the spinning particle model admitting
self-consistent interactions with external fields \cite{c16}.}
\begin{equation}
\label{a4}s=(\stackrel{.}{x}^2)^{1/2}\quad ,\qquad \xi =\frac{(\stackrel{.}{
\lambda }_a\stackrel{.}{x}^{ab}\lambda _b)(\stackrel{.}{\overline{\lambda }}_c
\stackrel{.}{x}^{cd}\overline{\lambda }_d)}{\stackrel{.}{x}^2(\overline{
\lambda }_a\stackrel{.}{x}^{ab}\lambda _b)^2}\quad ,\qquad \eta =
\frac{\epsilon ^{abcd}
\stackrel{.}{\lambda }_a\overline{\lambda }_b\stackrel{.}{\overline{\lambda }%
}_c\lambda _d}{(\overline{\lambda }_a\stackrel{.}{x}^{ab}\lambda _b)^2}\ ,
\end{equation}
where $\epsilon ^{abcd}$ is the Lorentz invariant spin-tensor \cite{c17},
totally antisymmetric in its indices. Then the most general Poincar\'e and
reparametrization invariant Lagrangian looks like: ${\cal L}=sF\left( \xi
,\eta \right) $, with $F$ being arbitrary function. To specify the
particular form of the Lagrangian we require the presence of three local
symmetries corresponding to the off-shell conservation of the N\"oether
charges associated with the classical counterparts of Casimir operators
\begin{equation}
\label{a5}
\begin{array}{c}
C_1=P^2+m^2\equiv 0 \\
\\
C_2=W_{ABC}W^{ABC}-m^2(\delta _1^2+\delta _2^2)\equiv 0\,,\quad
C_3=W_AW^A+m^2\delta _1^2\delta _2^2\equiv 0
\end{array}
\end{equation}
Here $W^A=\epsilon ^{ABCDEF}J_{BC}J_{DE}P_F$ , $W^{ABC}=\epsilon
^{ABCDEF}J_{DE}P_F$ are Pauli-Lubanski vector and tensor respectively; $%
P_A=p_A$ and $J_{AB}=$.$x_Ap_B-x_Bp_A+\left( \left( \sigma _{AB}\right)
_a\!^b\lambda _b\pi ^a+c.c.\right) $ are the N\"oether charges associated
with the global Poincar\'e invariance of the theory and, finally,
\begin{equation}
\label{a6}p_A=\frac{\partial {\cal L}}{\partial \stackrel{.}{x}^A}\quad
,\qquad \pi ^a=\frac{\partial {\cal L}}{\partial \stackrel{.}{\lambda }_a}
\end{equation}
are the canonical momenta. The parameter $m$ entering rels. (\ref{a5}) is
nothing but the mass of the particle while the parameters $\delta _1$ and $%
\delta _2$ relate to the particle's spin. The substitution of the explicit
expressions for the momenta (\ref{a6}) into the conditions (\ref{a5}) yields
the set of differential equations for $F,$ resolving which we come to the
following Lagrangian:
\begin{equation}
\label{a7}\displaystyle{\!\!{\cal L}=\!\!\sqrt{-\stackrel{.}{x}^2\!\!\left( m^2-4\delta
_1^2\frac{\epsilon ^{abcd}\stackrel{.}{\lambda }_a\overline{\lambda }_b%
\stackrel{.}{\overline{\lambda }}_c\lambda _d}{(\overline{\lambda }_a%
\stackrel{.}{x}^{ab}\lambda _b)^2}+4m\sqrt{\left( \delta _2^2-\delta
_1^2\right) \frac{\epsilon ^{abcd}\stackrel{.}{\lambda }_a\overline{\lambda }%
_b\stackrel{.}{\overline{\lambda }}_c\lambda _d}{(\overline{\lambda }_a%
\stackrel{.}{x}^{ab}\lambda _b)^2}}\right) }\!+\!2\delta _1\left| \frac{%
\stackrel{.}{\lambda }_a\stackrel{.}{x}^{ab}\lambda _b}{\overline{\lambda }_a%
\stackrel{.}{x}^{ab}\lambda _b}\right| }
\end{equation}
The Lagrangian (\ref{a7}) is obviously Poincar\'e invariant and possesses,
by construction, five gauge symmetries two of which are the local $\lambda $%
-rescalings: $\lambda _a\sim \alpha \lambda _a$ and one may be associated
with the reparametrizations of the particle's world-line.

In the Hamiltonian formalism the model is completely characterized by the
set of five first-class constraints three of which are dynamical
\begin{equation}
\label{a8}T_1=p^2+m^2\approx 0\,,\quad T_2=\overline{\lambda }%
_ap^{ab}\lambda _b\overline{\pi }^cp_{cd}\pi ^d+m^2\delta _1^2\approx 0
\,,\quad
T_3=\lambda _a\overline{\lambda }_b\pi ^a\overline{\pi }\!^b+\delta
_2^2\approx 0
\end{equation}
and the other two are kinematical
\begin{equation}
\label{a9}T_4=\pi ^a\lambda _a\approx 0\quad ,\qquad T_5=\overline{\pi }^a%
\overline{\lambda }_a\approx 0
\end{equation}
generating the $\lambda $-rescalings with respect to the canonical Poisson
brackets
\begin{equation}
\label{a10}\left\{ x^A,p_B\right\} =\delta ^A\!_B\,,\quad \left\{ \lambda
_a,\pi ^b\right\} =\delta ^b\!_a\,,\quad \left\{\overline{\lambda }_a,\overline{%
\pi }^b\right\}=\delta ^b\!_a
\end{equation}
The corresponding first-order (Hamiltonian) action associated with the
constraints (\ref{a8}), (\ref{a9}) looks like
\begin{equation}
\label{a11}S_H=\int d\tau \left\{ p_A\stackrel{.}{x}^A+\pi ^a\stackrel{.}{%
\lambda }_a+\overline{\pi }^a\stackrel{.}{\overline{\lambda }}%
_a-\sum\limits_{i=1}^5e_iT^i\right\} \ ,
\end{equation}
where $e_i$ are the Lagrange multipliers to the constraints, and $e_4=%
\overline{e}_5$. Notice that despite the quite nonlinear structure of the
constraints, the classical equations of motion are completely integrable for
the action (\ref{a11}) with the arbitrary Lagrange multipliers $e_i$. This
fact is not surprising as the model, by construction, describes a free
relativistic particle possessing sufficient number of symmetries. Here,
however, we omit the explicit expressions for the Hamiltonian equations as
well as its solution in the spinning sector. (The more detailed treatment of
this subject will be given in ref. \cite{c19}) In the Minkowski space the
corresponding solution reads
\begin{equation}
\label{a12}
\begin{array}{c}
x^A\left( \tau \right) =x_0^A+2\left( E_1+E_2\delta _1^2\right)
p^A-m^{-2}V^A\left( \tau \right) \\
\\
V^A\left( \tau \right) =V_1^A\cos \left( 2m^2E_2\delta _1\right) +V_2^A\sin
\left( 2m^2E_2\delta _1\right)
\end{array}
\end{equation}
Here $E_i\left( \tau \right) =\int\limits_0^\tau d\tau e_i(\tau )$, the
constant vector of the six-momentum $p_A$ is assumed to be chosen on the
mass shell and the constant vectors $V_1,V_2$, being defined by initial data
for the spinning degrees of freedom, are constrained to satisfy
\begin{equation}
\label{a13}p_AV_{1,2}^A=0\;,\quad V_1{}^2=V_2{}^2=\delta _1^2-\delta
_2^2\,,\quad (V_1,V_2)=0
\end{equation}
As is seen from (\ref{a12}) the space-time evolution is completely
determined by the independent evolution of the two Lagrange multipliers $%
e_1,e_3$. The presence of the additional gauge invariance in the solutions (%
\ref{a12}), as compared with the spinless particle case, causes the
conventional notion of the particle's world-line to fail. Instead, according
to (\ref{a12}), one has to consider the classes of gauge equivalent
trajectories which in the case under consideration are identified with the
two-dimensional tubes of radius $\rho =\sqrt{\delta _1^2-\delta _2^2}$ along
the particle's momenta $p_A$. So, in each moment of time (which may be
chosen by imposing the gauge fixing condition $x^0=c\tau $ ) the massive
spinning particle is not localized in a certain point of Minkowski space but
represents a string-like configuration contracting to a point only provided
that $\delta _1=\delta _2$.

Finally, let us discuss the structure of the physical observables of the
theory. Each physical observable $A$ being a gauge-invariant function on the
phase space should meet the requirements:
\begin{equation}
\label{a25}\left\{ A,T_i\right\} =0\quad ,\qquad i,j=1,..,5
\end{equation}
Due to the obvious Poincar\'e invariance of the constraint surface, the
generators $J_{AB},P_C$ automatically satisfy (\ref{a25}) and thereby are
observable. On the other hand, it is easy to compute that the dimensionality
equals 18 of the physical phase space of the theory. Thus the physical
subspace may covariantly be parametrized by 21 Poincar\'e generator subject
to 3 conditions (\ref{a5}) and as a result any physical observable proves to
be a function of the generators modulo constraints. So, a general solution
to (\ref{a25}) reads
\begin{equation}
\label{a26}A=f\left( J_{AB},P_C\right) +\sum\limits_{i=1}^5\alpha _iT_i\ ,
\end{equation}
$\alpha _i$ being arbitrary function of the phase space variables. In fact,
this implies that the physical phase space of the model is embedded in the
linear space of the Poincar\'e algebra through the constraints (\ref{a5})
and therefore coincides with the join of two 18-dimensional coadjoint orbits
of the Poincar\'e group. One of the orbits is associated with the particle
of positive energy $p_0>0$ and another corresponds to $p_0<0$.

\section{Quantization and relativistic harmonic analysis on $CP^3$}

As it was argued in previous section, the model is completely characterized
at the classical level by the algebra of observables associated with the
phase space generators of the Poincar\'e group, so that any gauge invariant
phase space function could be expressed via $J_{AB}$ and $P_C$. From this
point of view the quantization of the theory reduces to an explicit
construction of irreducible unitary representation of the Poincar\'e group
and may be carried out by means of geometrical quantization method \cite{c18}%
.

Within the framework of covariant operatorial quantization, the Hilbert
space of physical states of the system is embedded into the space of smooth
complex functions on $M^{12}$ and the phase space variables $x^A,p_A,\lambda
_a,\pi ^a$ are considered to be Hermitian operators subject to the canonical
commutation relations.

In the ordinary coordinate representation: $p_A\rightarrow -i\partial _A\ ,\
\pi ^a\rightarrow -i\partial ^a\ ,\ \overline{\pi }^a\rightarrow -i\overline{%
\partial }^a$ the quantum first-class constraints take the form
\begin{equation}
\label{a14}
\begin{array}{c}
\widehat{T}_1=\Box -m^2\,,\quad \widehat{T}_2=\partial^{ab}\overline{\lambda }%
_a\lambda _b\partial_{cd}\overline{\partial }^c\partial ^d+m^2\delta _1^2\,,\quad
\widehat{T}_3=\lambda _a\overline{\lambda }_b\overline{\partial }^a\partial
^b-\delta _2^2 \\  \\
\widehat{T}_4=-i\lambda _a\partial ^a\quad ,\qquad \widehat{T}_5=-i\overline{%
\lambda }_a\overline{\partial }^a
\end{array}
\end{equation}
The subspace of physical states is extracted by the conditions
\begin{equation}
\label{a15}\widehat{T}_i\left| \Phi _{phys}\right\rangle =0\quad ,\qquad
i=1,...,5
\end{equation}

Notice that the classical dynamics is consistent with arbitrary values of
the parameters $\delta _1,\delta _2$. Nevertheless, the nontrivial solutions
to the equations for physical states (\ref{a15}) can exist only provided
that $\delta _1^2=s_1\left( s_1+3\right) $ , $\delta _2^2=s_2\left(
s_2+1\right) $ where $s_1\geq s_2$ , $s_1,s_2=0,1,2,...$ .

Let us now consider the space $^{\uparrow }{\cal H}(M^{12},m)$ of massive
positive frequency fields on $M^{12}$. These fields are annihilated by the
constraints $\widehat{T}_1,\widehat{T}_4,\widehat{T}_5$ and possess the
Fourier decomposition
\begin{equation}
\label{a16}
\begin{array}{c}
\displaystyle{\Phi \left( x,\lambda ,\overline{\lambda }\right) =\int \frac{d%
\stackrel{\rightarrow }{p}}{p_0}e^{i(p,x)}\Phi \left( p,\lambda ,\overline{%
\lambda }\right) } \\  \\
p^2+m^2=0\quad ,\quad p_0>0
\end{array}
\end{equation}

The space $^{\uparrow }{\cal H}(M^{12},m)$ may be endowed with the
Poincar\'e-invariant and positive-definite inner product defined by the rule
\begin{equation}
\label{a17}\langle \Phi _1\left| \Phi _2\right\rangle =i\int \frac{d%
\stackrel{\rightarrow }{p}}{p_0}\int\limits_{CP^3}\omega \wedge \overline{%
\omega }\Phi _1\overline{\Phi }_2
\end{equation}
where the three-form $\omega $ is given by
\begin{equation}
\label{a18}\omega =\frac{\epsilon ^{abcd}\lambda _ad\lambda _b\wedge
d\lambda _c\wedge d\lambda _d}{\left( \overline{\lambda }_ap^{ab}\lambda
_b\right) ^2}
\end{equation}
Then $^{\uparrow }{\cal H}(M^{12},m)$ becomes the Hilbert space and, as a
result, the Poincar\'e group representation is unitary in this space. This
representation is decomposed into the direct sum of irreducible ones
\begin{equation}
\label{a19}^{\uparrow }{\cal H}(M^{12},m)=\bigoplus\limits_{\stackrel{%
\scriptstyle{s_1,s_2=0,1,2,...}}{s_1\geq s_2}}{}^{\uparrow }{\cal H}%
_{s_1,s_2}(M^{12},m)\ ,
\end{equation}
where the invariant subspace $^{\uparrow }{\cal H}_{s_1,s_2}(M^{12},m)$
realizes the Poincar\'e representation of mass $m$ and spin $\left(
s_1,s_2\right) $ and thereby satisfies all the quantum conditions (\ref{a15}%
). The explicit expression for an arbitrary field from $^{\uparrow }{\cal H}%
_{s_1,s_2}(M^{12},m)$ reads
\begin{equation}
\label{a20}\Phi \left( p,\lambda ,\overline{\lambda }\right) =\Phi \left(
p\right) ^{a_1...a_{s_1+s_2}b_1...b_{s_1-s_2}}\frac{\lambda _{a_1}...\lambda
_{a_{s_1}}\overline{\lambda }_{a_{s_1+1}}..\overline{\lambda }_{a_{s_1+s_2}}%
\overline{\lambda }_{b_1}...\overline{\lambda }_{b_{s_1-s_2}}}{\left(
\overline{\lambda }_ap^{ab}\lambda _b\right) ^{s_1}}
\end{equation}
Here the spin-tensor $\Phi \left( p\right)
^{a_1...a_{s_1+s_2}b_1...b_{s_1-s_2}}$ is considered to be $p$-transversal
\begin{equation}
\label{a21}p_{a_1b_1}\Phi \left( p\right)
^{a_1...a_{s_1+s_2}b_1...b_{s_1-s_2}}=0
\end{equation}
(for $s_1\neq s_2$) and its symmetry properties are described by the
following Young tableaux:

\unitlength=0.7mm
\special{em:linewidth 0.4pt}
\linethickness{0.4pt}
\begin{picture}(120.00,23.00)(00.00,122.00)
\put(30.00,140.00){\line(1,0){64.00}}
\put(94.00,140.00){\line(0,-1){8.00}}
\put(94.00,132.00){\line(-1,0){64.00}}
\put(30.00,132.00){\line(0,1){8.00}}
\put(30.00,132.00){\line(0,-1){8.00}}
\put(30.00,124.00){\line(1,0){32.00}}
\put(62.00,124.00){\line(0,1){8.00}}
\put(38.00,140.00){\line(0,-1){16.00}}
\put(54.00,140.00){\line(0,-1){16.00}}
\put(62.00,140.00){\line(0,-1){8.00}}
\put(70.00,140.00){\line(0,-1){8.00}}
\put(86.00,140.00){\line(0,-1){8.00}}
\put(34.00,136.00){\makebox(0,0)[cc]{$a_1$}}
\put(34.00,128.33){\makebox(0,0)[cc]{$b_1$}}
\put(45.80,135.92){\makebox(0,0)[cc]{. . .}}
\put(45.80,127.89){\makebox(0,0)[cc]{. . .}}
\put(77.94,136.10){\makebox(0,0)[cc]{. . . }}
\put(57.94,136.10){\makebox(0,0)[cc]{$a_n$}}
\put(57.94,128.07){\makebox(0,0)[cc]{$b_n$}}
\put(89.90,136.10){\makebox(0,0)[cc]{$a_m$}}
\put(120.00,136.00){\makebox(0,0)[cc]{$n=s_1-s_2$}}
\put(120.00,128.00){\makebox(0,0)[cc]{$m=s_1+s_2$}}
\end{picture}

The field $\Phi \left( p\right) ^{a_1...a_{s_1+s_2}b_1...b_{s_1-s_2}}$ can
be identified with the Fourier transform of spin-tensor field on Minkowski
space $R^{5,1}$. The mass-shell condition
\begin{equation}
\label{a22}\left( p^2+m^2\right) \Phi \left( p\right)
^{a_1...a_{s_1+s_2}b_1...b_{s_1-s_2}}=0
\end{equation}
and relation (\ref{a21}) constitute, together, the full set of relativistic
wave equations for the mass-$m$, spin-$\left( s_1,s_2\right) $ field in six
dimensions. Thus a massive scalar field on $M^{12}$ generates massive fields
of arbitrary integer spins in Minkowski space.

It is instructive to rewrite the inner product for two fields from $%
^{\uparrow }{\cal H}_{s_1,s_2}(M^{12},m)$ in terms of spin-tensors $\Phi
\left( p\right) ^{a_1...a_{s_1+s_2}b_1...b_{s_1-s_2}}$.The integration over
the spinning variables, being performed with the use of the integral $%
\int_{CP^3}\omega \wedge \overline{\omega }=-48i\pi ^3(p^2)^{-2}$, results
with
\begin{equation}
\label{a23}\langle \Phi _1\left| \Phi _2\right\rangle =N\int \frac{d%
\stackrel{\rightarrow }{p}}{p_0}\Phi _1\left( p\right) ^{a_1...a_{2s_1}}%
\overline{\Phi }_2\left( p\right) _{a_1...a_{2s_1}}\ ,
\end{equation}
where
\begin{equation}
\label{a24}\overline{\Phi }_2\left( p\right) _{a_1...a_mb_1...b_n}=\epsilon
_{a_1b_1c_1d_1}...\epsilon _{a_nb_nc_nd_n}p_{a_{n+1}c_{n+1}}...p_{a_mc_m}%
\overline{\Phi }_2\left( p\right) ^{c_1...c_md_1...d_n}
\end{equation}
and $N$ is some normalization constant depending on $s_1$ and $s_2$.

Notice that in the rest reference system $p_A=(m,0,...,0)$ the differential
operators entering the constraints $\widehat{T}_2,\widehat{T}_3$ turn to the
conventional $SO(5)$-invariant Laplace operators on $CP^3$ and rel. (\ref
{a19}) becomes the standard expansion for a scalar field on $CP^3$ via the
'spherical' functions. That is why the Poincar\'e invariant constructions of
this Section may be thought of as the {\it relativistic} harmonic analysis
on $CP^3$.

The treatment of the half-integer spin representations may be performed
along the similar lines by considering the space of special smooth tensor
fields on $M^{12}$ instead of the space of scalar functions $^{\uparrow }%
{\cal H}(M^{12},m)$ (cf. see \cite{c19}).

\section{Concluding remarks}

In this talk we have suggested the model for a massive spinning particle in
the six-dimensional Minkowski space as a mechanical system with
configuration space $M^{12}=R^{5,1}\times CP^3$. The Lagrangian of the model
is unambiguously constructed from the $M^{12}$ world line invariants when
the identical conservation is required for the classical counterparts of
Casimir operators.

Notice that switching on an interaction of the particle to the inhomogeneous
external field, one destroys the first class constraints algebra of the
model and the theory, thereby, becomes inconsistent, whereas the homogeneous
background is admissible. The physical cause underlying this inconsistency
is probably that the local nature of the inhomogeneous field may contradict
to the nonlocal behavior of the particle dynamical histories. The possible
method to overcome the obstruction to the interaction is to involve the
Wess-Zumino like invariants omitted in the action (\ref{a7}). The similar
trick solves this problem in the case of $d=4$ spinning particle \cite{c16}.

\section{Acknowledgments}

The authors would like to thank I. A. Batalin, I. V. Gorbunov, S. M.
Kuzenko, A. Yu. Segal and M. A. Vasiliev for useful discussions on various
topics related to the presented research. The work is partially supported by
the European Union Commission under the grant INTAS 93-2058. S. L. L. is
supported in part by the grant RBRF 96-01-00482.


\begin{thebibliography}{99}
\bibitem{c1}  Frenkel J., Z. f\"ur Physik {\bf 37} (1926), 243.

\bibitem{c2}  I. E. Tamm, V. L. Ginzburg, JETP {\bf 17} (1947), 227.

\bibitem{c3}  Barut A. O., ''{\it Electrodynamics and Classical Theory of
Fields and Particles}'', MacMillan New York 1964.

\bibitem{c4}  Frydryszak A., ''{\it Lagrangian models of particles with
spin: the first seventy years}'', hep-th/9601020.

\bibitem{c5}  Berezin F. A., Marinov M. S., Ann. Phys. {\bf 104} (1977) 336.

\bibitem{c6}  A. Barducci, R. Casalbuoni and L. Lusanna, Nuovo Cim. {\bf A35%
}, 377 (1976).

\bibitem{c7}  L. Brink, S. Deser, B. Zumino, P. Di Veccia and P. Howe, Phys.
Lett. {\bf B64} (1976) 435.

\bibitem{c8}  V. D. Gershun and V. I. Tkach, JETP Lett., {\bf 29} (1979) 288.

\bibitem{c9}  P. Howe, S. Penati, M. Pernicci and P. Townsend, Phys. Lett.
{\bf 215} (1988) 555;Class. Quantum Grav. {\bf 6} (1988) 1125.

\bibitem{c10}  A. O. Barut and N. Zanghi, Phys. Rev. Lett. {\bf 52}, 2009
(1984).

\bibitem{c11}  A. P. Balachandran, G. Marmo, A. Stern, and B.-S. Skagerstam,
{\it Gauge Symmetries and Fibre Bundles: Application to Particle Dynamics},
Lecture Notes in Physics, vol.188 (Springer, 1983).

\bibitem{c12}  R. Marnellius U. M\aa rtensson, Nucl. Phys. {\bf B335}, 395
(1991); Int. J.Mod. Phys. {\bf A6}, 807 (1991).

\bibitem{c13}  S. M. Kuzenko, S. L. Lyakhovich and A. Yu. Segal, Int. J.
Mod. Phys. {\bf A10}, 1529 (1995).

\bibitem{c14}  S. M. Kuzenko, S. L. Lyakhovich and A. Yu. Segal, Phys. Lett.
{\bf B348}, 421 (1995).

\bibitem{c15}  S. M. Kuzenko, S.L.Lyakhovich, A. Yu. Segal and A. A.
Sharapov, ''{\it Massive spinning particle on anti-de Sitter space}'',
hep-th/9509062, to appear in Int. J. Mod. Phys.{\bf A} (1996).

\bibitem{c16}  S.L.Lyakhovich, A. Yu. Segal and A. A. Sharapov, ''{\it A
universal model of D=4 spinning particle}'', hep-th/9603174.

\bibitem{c17}  T. Kugo and P. K. Townsend, Nucl. Phys. {\bf B221} (1983) 357.

\bibitem{c18}  N. M. J. Woodhouse,{\it \ Geometric Quantization}, Oxford
Science Publications, Oxford 1991.

\bibitem{c19}  S. L. Lyakhovich, A. A. Sharapov, and K. M. Shekhter, {\it in
preparation.}
\end{thebibliography}
\end{document}